\documentclass[11pt,reqno,twoside]{article}
\usepackage{amsmath,amssymb,theorem,color}
\theorembodyfont{\itshape}
\newtheorem{thm}{Theorem}[section]
\newtheorem{lem}[thm]{Lemma}
\newtheorem{prop}[thm]{Proposition}
{\theorembodyfont{\upshape}

\newtheorem{rem}[thm]{Remark}

}

\newcommand{\Proof}[1][]{\noindent{\itshape Proof#1. }}
\newcommand{\EndProof}{~$\Box$\bigskip}
\makeatletter
    
    \@addtoreset{equation}{section}
\makeatother

\def\mbR{{\mathbb R}}
\def\mbC{{\mathbb C}}
\def\mbE{{\mathbb E}}
\def\mbT{{\mathbb T}}
\def\mbZ{{\mathbb Z}}

\def\mcF{\mathcal{F}}

\def\mcH{\mathcal{H}}

\def\mcX{\mathcal{X}}

\def\ol{\overline}

\def\sm{\setminus}
\def\sbs{\subset}

\def\ssbs{\subset\subset}
\def\ptl{\partial}

\def\wt{\widetilde}
\def\wh{\widehat}
\def\notni{{\ni \!\!\!\!\sm}\,}
\def\notnii{{\ni \!\!\!\!\!\sm}\,}
\def\tn{|\text{\hskip -0.05 true cm}|\text{\hskip -0.05 true cm}|}
\def\d{\delta}    
\def\a{\alpha}    \def\b{\beta}              \def\d{\delta}
\def\D{\Delta}    \def\e{\varepsilon}        
  \def\g{\gamma}      \def\G{\Gamma}     
\def\l{\lambda}   \def\L{\Lambda}     \def\m{\mu}        \def\n{\nu}
\def\r{\rho}               
\def\p{\pi}      \def\P{\Pi}             
     \def\t{\tau}     
\def\x{\xi}       
\def\z{\zeta}           

\begin{document}
\title{Glauber and Kawasaki Dynamics for Determinantal Point Processes in Discrete Spaces}
\author{Myeongju Chae\footnote{Department of Applied Mathematics,
Hankyong National University, 67 Seokjeong-dong, Anseong-si,
Gyeonggi-do 456-749, Korea. E-mail: mchae@hknu.ac.kr} \, and\,  Hyun Jae Yoo\footnote{Department of Applied Mathematics,
Hankyong National University, 67 Seokjeong-dong, Anseong-si,
Gyeonggi-do 456-749, Korea. E-mail: yoohj@hknu.ac.kr} \footnote{Corresponding author}}
\date{}
  \maketitle

\begin{abstract}
We construct the equilibrium Glauber and Kawasaki dynamics on
discrete spaces which leave invariant certain determinantal point
processes. We will construct Fellerian Markov processes with
specified core for the generators. Further, we discuss the
ergodicity of the processes.
\end{abstract}
\noindent {\bf Keywords}. {Glauber and Kawasaki dynamics, determinantal
point processes, Papangelou intensity, invariant measure, ergodic process.}\\
{\bf Running head}. {Glauber and Kawasaki dynamics}\\
{\bf 2000 Mathematics Subject Classification}. 60J75, 60J80, 60K35,
82C20.


\section{Introduction}\label{sec:introduction}
In this paper we construct the equilibrium Glauber and Kawasaki
dynamics on the discrete particle systems such that certain
determinantal point processes are invariant under the Markov
processes.

The determinantal point processes, or fermion point processes, are
point processes whose correlation functions are given by
determinants of kernel operators. It was invented by Macchi \cite{M}
and then has been extensively investigated by many people. It
appears in many fields in mathematics and physics, for example, in
random matrix theory and in fermion particle systems. For the details we refer to \cite{Ly, ST1,
ST2, S} and references therein.

The construction of equilibrium
dynamics  is one of interesting subjects for the determinantal point processes.
One approach is to construct the  diffusion processes via Dirichlet forms \cite{MR}. During the last two decades there
have been many works on the construction of equilibrium diffusion processes for Gibbs measures
of interacting particle systems.
We notice that the interacting Brownian
particles of logarithmic potential, which is related to the
determinantal point process of sine kernel and is called Dyson's
model, was studied by Spohn \cite{Sp}. Some related works to Dyson's
model were also done recently in \cite{KT, O}.
The diffusion process
via Dirichlet form for the determinantal point processes in the
non-percolating regime, or equivalently in the high temperature or
small activity regime, was constructed by the second named author
\cite{Y1}.
The other approach, in particular for particle systems,
is to construct the particle birth and death processes, so called
Glauber dynamics, and the particle jump processes, called
Kawasaki dynamics. The general interacting particle systems in
discrete model were developed in detail by Liggett \cite{Li}.
 The Glauber and Kawasaki dynamics for the continuum models have been
investigated in the literature \cite{G1, G2, HS, KL, KLR, P}. Among
these, Kondratiev \textit{et al.} established the Dirichlet form
approach for the construction of the equilibrium Glauber and
Kawasaki dynamics for continuum systems so that the Gibbs measures
for the system are invariant under the Markov processes \cite{KL,
KLR}. For instance, the standard superstable interaction of Ruelle
\cite{R} falls into their regime of applications. Lytvynov and
Ohlerich applied the methods developed in \cite{KL, KLR} to
construct the equilibrium Glauber and Kawasaki dynamics that leave
invariant certain determinantal point processes in continuum model
\cite{LO}. The Glauber dynamics for discrete determinantal point
processes was studied by Shirai and the second named author
\cite{SY}. This paper can be regarded as a continuation of the work
in \cite{SY}. We emphasize that the Markov processes constructed in
this paper are Feller processes. That is, the semigroup acting on a
continuous function gives rise to another continuous function. But
the Markov processes constructed via Dirichlet forms are Hunt
processes and the semigroups act on $L^2$-functions.

Let us briefly sketch the contents of this paper. We consider the
infinite particle system, particles living on a discrete set, say
$E$. We consider the determinantal point process $\m$ on the
configuration space $\mcX$. The point process $\m$ has a defining
kernel operator $K$ of the type $K:=A(I+A)^{-1}$, where $A$ is a
positive definite, bounded linear operator on $l^2(E)$ that
satisfies some hypothesis (H) (see Section 2). Under the hypothesis
(H), $\m$ is known to be Gibbsian \cite{Y2}. The generators of
Glauber and Kawasaki dynamics have the form (see subsection 2.1 for
the details):
\[
L^{(\text{G})}f(\xi):= \sum_{x \in \xi}{d(x;\xi)[f(\xi \sm
x)-f(\xi)]}+\sum_{y \in E\sm \xi}{b(y;\xi)[f(y\xi) -f(\xi)]};
\]
and
\[ L^{(\text{K})}f(\xi):= \sum_{x \in \xi,y \in E\sm
\xi}{c(x,y;\xi)[f(y\xi \sm x)-f(\xi)]},
\]
where $d(x;\x)$ and $b(y;\x)$ are death and birth rates for Glauber
dynamics, and $c(x,y;\x)$ are the jump rates for Kawasaki dynamics.
In Subsection \ref{subsec:strategy}, following Liggett \cite{Li}, we 
introduce the basic strategy to show the existence and ergodicity of the dynamics. 
Next for point processes with Papangelou intensities, 
we will give necessary conditions for the rate functions to satisfy 
the detailed balance condition (Theorem \ref{thm:thm3.2}).
Then we apply these ideas to our model. Given a DPP $\m$ for a kernel 
operator which satisfies certain hypothesis, we give
some concrete formulas for the rate functions (Proposition \ref{prop:prop3.3}) 
and present the conditions for existence in the 
language of Papangelou intensities (Proposition \ref{prop:prop3.7}).  
To guarantee that the conditions for the existence and ergodicity
are satisfied, we need further stronger conditions introduced in
Assumption (A) in Section 4.
 Under the assumption (A), we finally
 construct the dynamics of our purpose (Theorem
\ref{thm:thm4.2}). In the Appendix, we provide with a proof of a
technical lemma, Lemma \ref{lem:lem4.1}, which is worth to be
noticed in itself.

Finally, comparing to \cite{SY}, we would like to mention that in
this paper the Kawasaki dynamics is included and the concept of
Papangelou intensities is used  in a crucial way to construct the
dynamics. After denoting the detailed balance condition by
Papangelou intensities, it is possible to choose the flip or jump
rates for the equilibrium dynamics in many ways. In \cite{SY}, we
dealt with just one choice among them (see Remark \ref{rem:rem3.4}).
However, at the moment, we have to assume seemingly almost the same
conditions as in \cite{SY}. Our future studies are addressed to
\begin{enumerate}
\item[-] the  construction of  the Glauber and Kawasaki dynamics under weaker
conditions, at least under the hypothesis (H);\\[-4ex]
\item[-] the  investigation of  the spectral gap, or log-Sobolev
inequalities for the generators of the Markov processes.
\end{enumerate}

\section{Preliminaries}\label{sec:preliminaries}
In this section we briefly recall the definition of Glauber and
Kawasaki dynamics for spin systems, or equivalently lattice gases.
Then we introduce the determinantal point process in discrete
spaces and their Gibbsianness.
\subsection{Glauber and Kawasaki Dynamics for Lattice Gases}

Let $E$ be any countable set. We have in mind the system of the
lattice space with $E = Z^d$, the $d$-dimensional integer lattice,
but we deal with arbitrary discrete set $E$.
Let $\mcX$ be the set of all subsets $\xi \subset E$, called the
configurations. For any subset $\L \subset E$ and $ \xi \in \mcX$,
we denote by $\xi_\L$ the restriction of $\xi$ to $\L$:
\begin{equation}\label{eq:eq2.1}
\xi_\L:= \xi \cap \L.
\end{equation}
From now on if a subset $\D \subset E$ is finite, we denote it by
$\D \subset\subset E$. For each $\L \subset E$, let $\mcF_\L$ be the
$\sigma$-algebra on $\mcX$ generated by the functions $\xi \mapsto
|\xi_\D|$, $\D \subset\subset \L$, where $|\xi_\D|$ means the
cardinality of the set $|\xi_\D|$. Thus we get an increasing system
of $\sigma$-algebras $\{\mcF_\L\}_{\L\subset\subset E}$ and we let
$\mcF:= \mcF_E$. Any probability measure $\m$ on $(\mcX,\mcF)$ is
called a point process. We notice that the $\sigma$-algebra $\mcF$
on $\mcX$ can be understood as a Borel $\sigma$-algebra by a natural
identification of $\mcX$ with $\{0,1\}^E$ equipped with the product
topology. This identification is taken for granted in this paper,
and consequently we will also consider $\mcX$ as a topological
space. Notice in particular that $\mcX$ is a compact space.

Let $C(\mcX)$ denote the set of all continuous functions on $\mcX$
equipped with the sup-norm $\|\cdot\|$. For $f \in C(\mcX)$ and $x
\in E$, let
\begin{equation}\label{eq:eq2.2}
\D_{f}(x):=\sup\{|f(x\xi)-f(\xi)|:\xi \in \mcX\},
\end{equation}
where we have used a shorthand notation $x\xi := \{x\} \cup \xi$.
\noindent We define a subset of "smooth" functions $D(\mcX)$ \cite{Li} by
\begin{equation}\label{eq:eq2.3}
D(\mcX):= \{ f \in C(\mcX) : \tn f\tn:= \sum_{x \in E} \D_f(x) <
\infty \}.
\end{equation}
By the Stone-Weierstrass theorem, we easily see that $D(\mcX)$ is
dense in $C(\mcX)$.

The generators for Glauber and Kawasaki dynamics are defined as
follows. We first consider Glauber dynamics. For each $x \in E$, let
$b(x;\xi)$ and $d(x; \xi)$ be nonnegative continuous functions on
$\mcX$. They are called birth and death rates, respectively. Namely,
given a configuration $\x\in \mcX$, a particle is born at site
$x\notin \x$ in a rate $b(x;\x)$, and among the particles $\x$ a
particle at $x\in \x$ dies out in a rate $d(x;\x)$. For each $x \in
E$, we define
\begin{equation}\label{eq:eq2.4}
c_x := \sup _{\xi \notni x}\max\{b(x;\xi), d(x;x\xi)\} ,
\end{equation}
and define
\begin{equation}\label{eq:eq2.5}
c^{(\text{G})}:=\sup_{x \in E} c_x
\end{equation}
Next for Kawasaki dynamics, we need the particle jump rates. For
each $x\neq y\in E$, let $c(x,y;\x)$ be a continuous function on
$\mcX$ whose values are defined to be zero unless $x\in \x$ and
$y\notin \x$. It is the rate for a particle at $x\in \x$ to jump to
the empty site $y\in E\sm\x$. We will also need to control the
particle jump rates. For each pair $x \neq y \in E$, define
\begin{equation}\label{eq:eq2.6}
c_{\{x,y\}} := \sup_{\xi \notni x,y}\max \{c(x,y;x\xi), c(y, x ;y\xi)\};
\end{equation}
\begin{equation}\label{eq:eq2.7}
c^{(\text{K})}:= \sup_{x \in E} {\sum_{y \neq x}c_{\{x,y\}}}.
\end{equation}
We will assume that $c^{(\sharp)}< \infty$ for $\sharp=\text{G}$ or
$\text{K}$. Under this assumption, the Markov pregenerators for
Glauber and Kawasaki dynamics are defined as follows. For $f \in
D(\mcX)$, we define for Glauber dynamics
\begin{equation}\label{eq:eq2.8}
L^{(\text{G})}f(\xi):= \sum_{x \in \xi}{d(x;\xi)[f(\xi \sm
x)-f(\xi)]}+\sum_{y \in E\sm \xi}{b(y;\xi)[f(y\xi) -f(\xi)]},
\end{equation}
and for Kawasaki dynamics
\begin{equation}\label{eq:eq2.9}
L^{(\text{K})}f(\xi):= \sum_{x \in \xi,y \in E\sm
\xi}{c(x,y;\xi)[f(y\xi \sm x)-f(\xi)]}.
\end{equation}
Here we have also used a short notation $\xi \sm x$ for $\xi \sm
\{x\}$. For a definition and proof that $L^{(\sharp)}$,
$\sharp=\text{G or K}$, becomes a Markov pregenerator, we refer to
\cite[Definition 2.1 and Proposition 3.2]{Li}.

In this paper we will investigate the conditions on the rates
$b(x;\xi)$, $d(x;\xi)$, and $c(x,y;\xi)$ so that not only (the
closure of) $L^{(\sharp)}$, $\sharp=\text{G or K}$, becomes a
Markov generator on $C(\mcX)$ but also it leaves invariant certain
point process $\mu$ on $(\mcX,\mcF)$. We are concerned exclusively
with determinantal point proccesses on $E$, which are briefly
introduced in the next subsection.

\subsection{Determinantal Point Processes}
Determinantal point processes (DPP's hereafter), or fermion point
processes, are the probability measures on the configuration space
of particles. The particles can stay either on continuum spaces or
on discrete sets. The correlation functions of DPP's are given by
determinants of a priori given kernel operator as shown in Theorem
\ref{thm:thm2.1} below. Typically, they have a fermionic nature,
namely, the energy increases when new particles add into a given
configuration of particles. For the basic theories of DPP's we
refer to references \cite{{Ly},{M},{ST2},{S}}. Here we follow the
reference \cite{ST2} for the introduction.

Let $E$ be the countable set in the previous subsection and let $K$
be a Hermitian positive definite, bounded linear operator on the
Hilbert space $\mcH_0:=l^2(E)$, the set of square summable functions
(sequences) on $E$ equipped with the usual inner product:
\begin{equation}\label{eq:eq2.10}
(f,g)_0:= \sum_{x \in E} \ol{f(x)}g(x), \quad f,g \in \mcH_0.
\end{equation}
The induced norm is denoted by $\| \cdot \|_0$. The following is an
existence theorem for DPP's, which we present in the form stated in \cite{ST2}.
\begin{thm}\label{thm:thm2.1}
Let $E$ be a countable set  and  $K$ a Hermitian  bounded operator on $\mcH_0$.
Assume that $0 \leq K \leq I$. Then, there exists a unique probability measure
$\mu$ on $(\mcX,\mcF)$ such that for any finite set $\{ x_1,\ldots,x_n \} \sbs E$,
the $n$-th correlation function is given as follows:
\begin{equation}\label{eq:eq2.11}
\r_\mu^{(n)}(x_1,\ldots,x_n) := \mu(\{\xi \in \mcX : \xi \supset \{
x_1,\ldots,x_n \}\}) = \det (K(x_i,x_j))_{1\leq i,j \leq n}.
\end{equation}
\end{thm}

Next we discuss the density functions for DPP's. For each subset $\L \sbs E$, let $P_\L$
denote the projection operator on $\mcH_0$ onto the subspace $l^2(\L)$ and
let $ K_\L := P_\L K P_\L$ denote the restriction of $K$ onto the projection space.
For each $\L \ssbs E$, assuming at the moment that $I_\L-K_\L$ is invertible, we define
\begin{equation}\label{eq:eq2.12}
A_{[\L]}:=K_\L(I_\L-K_\L)^{-1}.
\end{equation}
The local marginals of the DPP $\mu$ corresponding to the operator
$K$ are given by the formula: for each $\L \ssbs E$ and fixed $\xi
\in \mcX$,
\begin{equation}\label{eq:eq2.13}
\mu_\L(\xi_\L):= \mu(\{\zeta : \zeta_\L= \xi_\L\})=\det(I_\L-K_\L)\det(A_{[\L]}(x,y))_{x,y \in \xi_\L},
\end{equation}
where $A_{[\L]}(x,y)$, $x,y \in \L$, denotes the matrix components
of $A_{[\L]}$. We remark that the r.h.s. of \eqref{eq:eq2.13} can be
given a meaning even when $I_\L-K_\L$ is not invertible \cite{ST2,
S}.
\subsection{Reproducing Kernel Hilbert Spaces, Papangelou Intensities and Gibbsianness of DPP's}

In this subsection we briefly discuss the Gibbsianness of DPP's. To
show the Gibbsianness of a point process $\mu$ is equivalent to show
the existence of global Papangelou intensities of $\m$ \cite{MWM,
NZ, Sh, ST2}. Since the Papangelou intensities are the key
ingredients for the construction of equilibrium dynamics we review
it for our model from \cite{Y2}. We start by introducing a dual pair
of reproducing kernel Hilbert spaces \cite{A}.

Let $A$ be a positive definite, bounded linear operator on $\mcH_0
\equiv l^2(E)$ (the operator will define a DPP via the operator
$K:=A(I+A)^{-1}$). We assume that $\text{Ker} A =\{0\}$, so
$\text{Ran} A$ is dense in $\mcH_0$. We introduce two new inner
products $(\cdot,\cdot)_-$ and $(\cdot,\cdot)_+$, respectively on
$\mcH_0$ and $\text{Ran} A$ as follows. First on $\mcH_0$, define
\begin{equation}\label{eq:eq2.14}
(f,g)_-:=(f,Ag)_0, \quad f,g \in \mcH_0,
\end{equation}
and on $\text{Ran}A$ define
\begin{equation}\label{eq:eq2.15}
(f,g)_+ := (f, A^{-1}g)_0, \quad f,g \in \text{Ran}A .
\end{equation}
The induced norms will be denoted by $\|\cdot\|_-$and $\|\cdot\|_+$,
respectively. Let $\mcH_-$ and $\mcH_+$ be the completions of
$\mcH_0$ and $\text{Ran}A$ by the respective norms $\|\cdot\|_-$and
$\|\cdot\|_+$. We then get the following  rigging of Hilbert spaces:
\begin{equation}\label{eq:eq2.16}
\mcH_- \supset \mcH_0 \supset \mcH_+ .
\end{equation}
We let ${\sf B}:=\{e_x\}_{x \in E}$ be the usual basis of $\mcH_0$,
i.e., $e_x$ is a function on $E$ whose value at $x$ is $1$ and the
values at other sites are all zero. Let $A(x,y), x,y \in E$, be the
representation of $A$ w.r.t. $\sf B$. Then we notice that $\mcH_+$ is a
reproducing kernel Hilbert space (RKHS hereafter) with reproducing
kernel (RK shortly) $A(x,y)$. That is, $\mcH_+$ is a linear space of
functions on $E$ with the following defining conditions:
\begin{enumerate}
\item[(i)] For every $x \in E$, the function $A(\cdot,x)$ belongs to
$\mcH_+$; \\[-4ex]
\item[(ii)] The reproducing property: for every $x \in E$ and $ g \in
\mcH_+$, $g(x)=(A(\cdot,x),g)_+$.
\end{enumerate}
We want $\mcH_-$ to be also a RKHS (it is not the case in general),
so we assume the hypothesis below throughout this paper:\\[2ex]
\noindent{\bf Hypothesis} (H): We suppose that $\mcH_-$ is functionally completed \cite{A},
i.e., any vector of $\mcH_-$ can be represented as a function on $E$. \\[2ex]
For sufficient conditions for the hypothesis (H) we refer to
\cite[Appendix]{Y2}. Under the hypothesis (H), $\mcH_-$ becomes also
a RKHS with RK, say $B(x,y), x,y \in E$ \cite{A, Y2}. Informally
speaking, $B(x,y)$ is the matrix representation of $A^{-1}$, though
it is not of bounded operator in general. In particular, under (H)
we have $e_x \in \mcH_+$ for all $x \in E$ \cite{Y2}. The flip or
jump rates of our dynamics will be defined via the Papangelou
intensities (defined below) of the DPP $\m$ corresponding to the
operator $K:=A(I+A)^{-1}$, which are in turn the squared norms of
the vector-projections in the Hilbert space $\mcH_-$  \cite{Y2}. In
particular, the interdependencies of rates, which must be controlled
for the construction of the dynamics, are represented by the inner
products in the restricted Hilbert spaces of $\mcH_-$ (see
Proposition \ref{prop:prop3.8}).

As mentioned above, we are concerned with DPP's corresponding to the
operators $K:=A(I+A)^{-1}$. In order to get the Papangelou
intensities, we need a variational principle \cite{Y2}. For each $\L
\ssbs E$, let
\begin{equation}\label{eq:eq2.17}
{\sf F}_{\text{loc},\L}:= \text{the linear space spanned by
}\{e_x: x \in \L\}.
\end{equation}
Let $x_0 \in E$ be a fixed point and let $ E= \{x_0\} \cup R_1 \cup
R_2$ be any partition of $E$ (one of $R_1$ and $R_2$  might be the
empty set). For each $\L \subset\subset E$, define
\begin{equation}\label{eq:eq2.18}
\alpha_{\L}:= \inf _{f\in {\sf F}_{\text{loc}, \L \cap R_1}}
\|e_{x_0} -f\|_-^2  \quad \text{and} \quad  \beta_\L :=  \inf _{ g
\in {\sf F}_{\text {loc}, \L \cap R_2}} \|e_{x_0} -g\|^2_+.
\end{equation}
Obviously, $ \{ \alpha_\L \}_{\L \subset\subset E}$ and $ \{
\beta_\L \}_{\L \subset\subset E}$  are decreasing nets of
nonnegative numbers. Consequently we define
\begin{equation}\label{eq:eq2.19}
\alpha := \lim_{\L \uparrow E}\alpha _{\L}\quad \text{and} \quad  \beta := \lim _{\L \uparrow E}\beta_{\L}.
\end{equation}
The variational principle says that, under the hypothesis (H), no matter how we take a
partition $E=\{x_0\} \cup R_1 \cup R_2$, the product of $\alpha$ and
$\beta$  is equal to 1 (see \cite[Theorem 2.4]{Y2} and also
\cite{ST2}):
\begin{equation}\label{eq:eq2.20}
\alpha\beta=1.
\end{equation}
The relation \eqref{eq:eq2.20} guaranties, on the other hand, the existence of global
Papangelou intensities.  Let $\mu$ be the DPP corresponding to the
operator $K:=A(I+A)^{-1}$, where $A$ satisfies the hypothesis (H).
Recall that the local Papangelou intensities of $\mu$ is defined as
follows: for each $x \in E$, $x \in \L \ssbs E$, and $x \notin \xi
\in \mcX$,
\begin{equation}\label{eq:eq2.21}
\alpha_{[\L]}(x; \xi_{\L}):=
\frac{\mu_{\L}(x\xi_{\L})}{\mu_{\L}(\xi_{\L})}.
\end{equation}
The global Papangelou intensities are the limits
\begin{equation}\label{eq:eq2.22}
\alpha(x;\xi) := \lim_{\L \uparrow E} \alpha_{[\L]}(x;\xi_{\L}),
\end{equation}
whenever the limit exists. The following theorem was proved in
\cite[Theorem 2.6]{Y2}:
\begin{thm}\label{thm:thm2.2}
Let the operator $A$ satisfy the hypothesis (H) and let $\mu$ be the
DPP corresponding to the operator $K=A(I+A)^{-1}$. Then for all $x
\in E$ and $x \notin \xi \in \mcX$, the Papangelou intensity
$\alpha(x;\xi)$ exists and it is equal to the the number $\alpha$ in
\eqref{eq:eq2.19} obtained by replacing $x_0$ and $R_1$ in
\eqref{eq:eq2.18} by $x$ and $\xi$, respectively.
\end{thm}
\begin{rem}\label{rem:gibbsian}
By \eqref{eq:eq2.13} and
\eqref{eq:eq2.21}-\eqref{eq:eq2.22} we see that
\begin{equation}\label{eq:eq3.1}
\a(x; \xi)=\lim_{\L \uparrow E}\frac{\det{
A_{[\L]}(x\xi_{\L},x\xi_{\L})}}{\det{ A_{[\L]}(\xi_{\L},\xi_{\L})}},
\end{equation}
here $A_{[\L]}(\xi_{\L},\xi_{\L})$ is the matrix
$(A_{[\L]}(x,y))_{x,y \in \xi_{\L}}$. What we have shown in the above theorem is that it is equal to the limit
\begin{equation}\label{eq:eq3.2}
\a(x; \xi)=\lim_{\L \uparrow E}\frac{\det{
A(x\xi_{\L},x\xi_{\L})}}{\det{ A(\xi_{\L},\xi_{\L})}}.
\end{equation} 
\end{rem}
We will denote the dual relation \eqref{eq:eq2.20} as
\begin{equation}\label{eq:eq2.23}
\a(x;\x)\b(x;\x)=1.
\end{equation}
We notice also that for all $x\in E$ and $x\notin \x\in \mcX$,
\begin{equation}\label{eq:eq2.24}
\a(x;\x)\le A(x,x)\le \|A\|.
\end{equation}
\section{Construction of Glauber and Kawasaki Dynamics}\label{sec:Construction of Glauber and Kawasaki Dynamics}
In this section we will construct the Glauber and Kawasaki dynamics
for DPP's. We begin by briefly introducing the general strategy for
the existence and ergodicity of the dynamics following \cite{Li}.
\subsection{Existence and Ergodicity}\label{subsec:strategy}
 Throughout this subsection we assume that the flip rates $b(x;\xi)$, $d(x;\xi)$, and
 jump rates $c(x,y;\xi)$ are continuous functions for $\x\in \mcX$ and satisfy the boundedness conditions $c^{(\#)}<\infty$, $\#=\text{G or
 K}$. Under these conditions the operators $L^{(\#)}$, $\#=\text{G or K}$,
in \eqref{eq:eq2.8}-\eqref{eq:eq2.9} are Markov pregenerators
\cite[Proposition 3.2, Chapter I]{Li}.

The Markov pregenerators are closable \cite[Proposition 2.5, Chapter
I]{Li}, but in order that the closures to be Markov generators, we
need to control the interdependencies of the rate functions. For Glauber
dynamics, following \cite{Li}, we define for each $x\neq u\in E$,
\begin{equation}\label{eq:eq3.9}
\g^{(\text{G})}(x,u):=\sup_{\x\notni
x,u}\{|b(x;\x)-b(x;u\x)|+|d(x;xu\x)-d(x;x\x)|\}.
\end{equation}
And we define
\begin{equation}\label{eq:eq3.10}
M^{(\text{G})}:=\sup_{x\in E}\sum_{u\neq x}\g^{(\text{G})}(x,u).
\end{equation}
We will need also the lower bound of the flip rates defined by
\begin{equation}\label{eq:eq3.11}
\e^{(\text{G})}:=\inf_{x\in E}\inf_{\x\notni x}\{b(x;\x)+d(x;x\x)\}.
\end{equation}
The interdependencies for Kawasaki dynamics will be given in the
following way. First define for each $x\neq u\in E$
\begin{equation}\label{eq:eq3.12}
\g^{(\text{K})}(x,u):=\sum_{y\neq x}\sup_{\x\notni
x,y,u}\{|c(x,y;x\x)-c(x,y;xu\x)|,\,\,|c(y,x;y\x)-c(y,x;yu\x)|\}.
\end{equation}
Then we define
\begin{equation}\label{eq:eq3.13}
M^{(\text{K})}:=\sup_{x\in E}\sum_{u\neq x}\g^{(\text{K})}(x,u).
\end{equation}
The lower bound of the rates for Kawasaki dynamics is given by
\begin{equation}\label{eq:eq3.14}
\e^{(\text{K})}:=\inf_{y\in E}\inf_{\x\notni y}\{\sum_{x\in
\x}c(x,y;\x)+\sum_{x\notin y\x}c(y,x;y\x)\}.
\end{equation}
Defining $\g^{(\#)}(x,x)\equiv
0$, we let $\G^{(\#)}=( \g^{(\#)}(x,u))_{x,u\in E}$ for $\#=\text{G or K}$. 
The following theorem is proven by
Liggett \cite[Theorem 3.9, Chapter I]{Li}:
\begin{thm}\label{thm:thm3.5}
For $\#=\text{G or K}$, assume that $c^{(\#)}<\infty$ and
$M^{(\#)}<\infty$. Then we have
\begin{enumerate}
\item[(a)] The closure $\ol{ L^{(\#)}}$ of $L^{(\#)}$ is a Markov
generator of a Markov semigroup $T_t^{(\#)}$.\\[-4ex]
\item[(b)] $D(\mcX)$ is a core for $\ol {L^{(\#)}}$.\\[-4ex]
\item[(c)] For $f\in D(\mcX)$
\[
\D_{T_t^{(\#)}f}\le e^{-\e^{(\#)}t}\exp(t\G^{(\#)})\D_f.
\]\\[-4ex]
\item[(d)] If $f\in D(\mcX)$, then $T_t^{(\#)}f\in D(\mcX)$ for all
$t\ge 0$ and
\[
\tn T_t^{(\#)}f\tn\le \exp[(M^{(\#)}-\e^{(\#)})t]\,\tn f\tn.
\]
\end{enumerate}
\end{thm}

Let us now introduce the concept of ergodicity. Recall that a point process $\n$ on $(\mcX,\mcF)$ 
is said to be invariant for the Markov process with semigroup $\{T_t,t\ge 0\}$ if 
\[
\int T_tfd\n=\int fd\n
\]
for all $f\in C(\mcX)$ and $t\ge 0$. A Markov
process with semigroup $\{T_t,t\ge 0\}$ is said to be ergodic
\cite{Li} if
\begin{enumerate}
\item[(i)] there is a unique invariant measure, say $\n_0$;\\[-4ex]
\item[(ii)] $\lim_{t\to \infty}\n T_t=\n_0$ for all probability
measure $\n$ on $(\mcX,\mcF)$.
\end{enumerate}
\begin{thm}\label{thm:thm3.6}
(\cite[Theorem 4.1, Chapter I]{Li}) Suppose that the same conditions
as in Theorem \ref{thm:thm3.5} hold. In addition, if
$M^{(\#)}<\e^{(\#)}$ then the process is ergodic. Furthermore, for
$g\in D(\mcX)$,
\[
\|T_t^{(\#)}g-\int g d\n_0^{(\#)}\|\le
c^{(\#)}\,\frac{e^{-(\e^{(\#)}-M^{(\#)})t}}{\e^{(\#)}-M^{(\#)}}\,\tn
g\tn,
\]
where $\n_0^{(\#)}$ is the unique invariant measure.
\end{thm}
\subsection{Detailed Balance Condition}\label{subsec:db}
In this subsection we discuss the detailed balance conditions for
Glauber and Kawasaki dynamics. Let $\m$ be a point process and
suppose that $\m$ has Papangelou intensities $\a(x;\x)$. By the
general theory, this implies that $\m$ satisfies the DLR-conditions,
which read as follows: for all bounded measurable functions $F:\mcX
\rightarrow \mbR$ and $\L \ssbs E$,
\begin{equation}\label{eq:eq3.4}
\int\mu(d\xi)F(\xi)
=\int\mu(d\xi)\frac{1}{Z_{\L}(\xi)}\sum_{\zeta_{\L} \sbs \L}
\a(\z_{\L};\xi_{\L^c})F(\zeta_{\L}\xi_{\L^c}),
\end{equation}
where $Z_{\L}(\xi)=\sum_{\zeta_{\L} \sbs \L} \a(\z_{\L};\xi_{\L^c})$
and $\a(\z_{\L};\xi_{\L^c}) =\prod_{i=1}^{|\z_{\L}|}\a(x_i;x_1\cdots
x_{i-1}\xi_{\L^c})$. We refer to \cite{MWM, NZ} for more details.

The detailed balance condition for Glauber and Kawasaki dynamics w.r.t. $\m$ 
means that the pregenerators $L^{(\sharp)}$, $\sharp=\text{G or K}$,
are symmetric:
\begin{equation}\label{eq:eq3.5}
\int\mu(d\xi)f(\xi)L^{(\sharp)}g(\xi)=\int\mu(d\xi)L^{(\sharp)}f(\xi)g(\xi),
\quad f,g \in D(\mcX).
\end{equation}
From the DLR-conditions \eqref{eq:eq3.4} it is not hard to get
equivalent conditions for the detailed balance. For simplicity, we
assume that the Papangelou intensities $\a(x;\x)$ are positive.
\begin{thm}\label{thm:thm3.2}
Let $L^{(\sharp)}$, $\sharp=\text{G or K}$, be the Markovian
pregenerators for Glauber and Kawasaki dynamics given in
\eqref{eq:eq2.8} and \eqref{eq:eq2.9}. In order that the detailed
balance condition \eqref{eq:eq3.5} is satisfied, it is necessary and
sufficient that the rate functions satisfy
\begin{enumerate}
\item[(a)] for Glauber dynamics: for all $x \in E$, $x \notin \xi \in
\mcX$,
\begin{equation}\label{eq:eq3.6}
b(x;\xi)=\a(x;\xi)d(x;x\xi);
\end{equation}\\[-4ex]
\item[(b)] for Kawasaki dynamics: $\forall x \neq y \in E$, and $\forall
\xi \in \mcX$ with $x, y \neq \xi$,
\begin{equation}\label{eq:eq3.7}
\a(x;\xi)c(x,y;x\xi)=\a(y;\xi)c(y,x;y\xi).
\end{equation}
\end{enumerate}
\end{thm}

\subsection{Glauber and Kawasaki Dynamics for
DPP's}\label{subsec:gkdyn_dpp} In this subsection, we will
concretely define the flip and jump rates for Glauber and Kawasaki
dynamics for DPP's, and then discuss the existence and ergodicity
conditions for the dynamics.

From now on we fix an operator $A$ on $\mcH_0$ that satisfies the
hypothesis (H). Thereby we also fix a DPP $\mu$ with kernel operator
$K:=A(I+A)^{-1}$. We will construct Fellerian Markov generators for
Glauber and Kawasaki dynamics on $C(\mcX)$ that leave invariant the
above DPP $\mu$. Recall the notations in subsection 2.1. In order
that the pregenerators (which are not yet explicitly defined)
$L^{(\text{G})}$ and $L^{(\text{K})}$ in \eqref{eq:eq2.8} and
\eqref{eq:eq2.9} would define Markovian generators for Feller
processes, it is needed that the functions $b(x;\xi)$, $d(x;\xi)$,
and $c(x,y;\xi)$ are continuous functions for $\xi \in \mcX$. These
flip, or jump rates will be given via the Papangelou intensity
function $\alpha(x;\xi)$. So, we need the following
\begin{lem}\label{lem:lem3.1}
For each $x \in E$, the Papangelou intensity $\alpha(x;\xi)$ of $\m$
is a continuous function of $\xi \in \mcX$.
\end{lem}
\Proof Recall the definitions $\alpha_{\L}$ and $\beta_{\L}$ in
\eqref{eq:eq2.18}. For each $x \in E$ and $\xi \in \mcX$ with $x
\notin \xi$, we consider the partition $E=\{x\}\cup \xi \cup
\ol{\xi}$, where $\ol{\xi}=E \sm (x\xi)$. For each $\L \ssbs E$ we
define local functions $\alpha_{\L}(x;\xi_{\L})$ and
$\beta_{\L}(x;\xi_{\L})$ by the formula \eqref{eq:eq2.18} replacing
$x_0$ and $R_1$ by $x$ and $\xi$, respectively.  As local functions,
clearly the functions $\xi \mapsto \a_{\L}(x;\xi_{\L})$ and $\xi
\mapsto \b_{\L}(x;\xi_{\L})$ are continuous on the set
$\{\xi \in \mcX : \xi \notnii x\}$. Now as decreasing limits of
continuous functions, the Papangelou intensity $\a(x;\xi)$ and
$\b(x;\xi)$ are both upper semi-continuous functions on $\{\xi \in
\mcX : \xi \notnii x\}$. Since $\a(x;\xi)$ and $\b(x;\xi)$ are
reciprocal to each other, they are also lower semi-continuous. This
proves the lemma. \EndProof

In the rest of the paper we will use the rate functions given in the
following proposition, which we can easily prove. Recall the variational relation
$\a(x;\x)\b(x;\x)=1$ for all $x\in E$ and $x\notin \x\in \mcX$ in \eqref{eq:eq2.23}.
\begin{prop}\label{prop:prop3.3}
Assume the hypothesis (H) and let $\m$ be the DPP corresponding to
the operator $K:=A(I+A)^{-1}$. Then the following choices for flip
rates $b(x;\x)$ and $d(x;\x)$ for Glauber dynamics, and jump rates
$c(x,y;\x)$ for Kawasaki dynamics are uniformly bounded for $\x \in
\mcX$ and satisfy the detailed balance conditions w.r.t. $\m$: (we
let $x, y \notin  \xi$)
\begin{enumerate}
\item[(a)] for Glauber dynamics:
\begin{equation}\label{eq:eq3.7-1}
b(x;\x):=\frac{\a(x;\x)}{1+\a(x;\x)}\quad\text{and} \quad
d(x;x\x):=\frac{\b(x;\x)}{1+\b(x;\x)};
\end{equation}\\[-4ex]
\item[(b)] for Kawasaki dynamics:
\begin{equation}\label{eq:eq3.7-2}
 c(x,y;x\xi)=d(x,y)\a(y;\x)g(\a(x;\x),\a(y;\x)),
 \end{equation}
 where $d(x,y)$ is a symmetric weight function, and $g:\mbR_+^2\to \mbR_+$ is
 any symmetric, continuous function that makes $c(x,y;x\x)$ bounded.
 \end{enumerate}
 \end{prop}
The most simplest example for the function $g$ in
the above is $g(u,v)\equiv 1$. However, reflecting on the
nature of the dynamics, including this case, we may choose for any
$0\le t\le 1$,
\begin{equation}\label{eq:eq3.7-3}
g(u,v)\equiv g_t(u,v):=\left(\frac{1}{(1+u)(1+v)}\right)^t.
\end{equation}
Then $c(x,y;x\x)$ becomes
\begin{equation}\label{eq:eq3.7-4}
c(x,y;x\x)=d(x,y)\b(x;\x)^t\a(y;\x)^{1-t}\left(\frac{1}{(1+\b(x;\x))(1+\b(y;\x))}\right)^t.
\end{equation}
\begin{rem}\label{rem:rem3.4}
In the equations \eqref{eq:eq3.7-1} and \eqref{eq:eq3.7-4}, the
terms $(1+\b(x;\x))^{-1}$ and $((1+\b(x;\x))(1+\b(y;\x)))^{-t}$ are
to make the flip rates bounded. When $\a(x;\x)$ is uniformly (for
$\x$) away from $0$, we may drop these terms (taking $b(x;\x)\equiv
\a(x;\x)$ and $d(x;\x)\equiv 1$ for Glauber dynamics). In \cite{SY},
we have taken, in our terminology, $b(x;\x)=1+\a(x;\x)$ and
$d(x;x\x)=1+\b(x;\x)$ (see \cite[eq. (1.9)]{SY}).
\end{rem}
Now the flip rates for the equilibrium Glauber and Kawasaki dynamics
have been given via the Papangelou intensities $\a(x;\x)$, we would like
to represent the condition $M^{(\#)}<\infty$ in terms of
$\a(x;\x)$. Let us assume that the weight
function $d(x,y)$ in \eqref{eq:eq3.7-2} satisfies
 \begin{equation}\label{eq:eq3.8}
 0<d_1:=\inf_{x\in E}\sum_{y\neq x}d(x,y)\le \sup_{x \in E}\sum_{y \neq x}d(x,y)=:d_2 <
 \infty.
 \end{equation}
\begin{prop}\label{prop:prop3.7}
Suppose that the flip rates $b(x;\x)$ and $d(x;\x)$ and the jump
rates $c(x,y;x\x)$ are given by the equations
\eqref{eq:eq3.7-1}-\eqref{eq:eq3.7-2} with the function $g$ in
\eqref{eq:eq3.7-2} given by \eqref{eq:eq3.7-3}, and $d(x,y)$
satisfying \eqref{eq:eq3.8}. Define
\begin{equation}\label{eq:eq3.17}
M_1^{(\text{G})}:=\sup_{x\in E}\sum_{u\neq x}\sup_{\x\notni
x,u}(\a(x;\x)-\a(x;u\x))
\end{equation}
and
\begin{equation}\label{eq:eq3.18}
M_1^{(\text{K})}:=\sup_{x\in E}\sum_{u\neq x}\sum_{y\neq
x}d(x,y)\sup_{\x\notni
x,y,u}\left[(\a(x;\x)-\a(x;u\x))+(\a(y;\x)-\a(y;u\x))\right].
\end{equation}
Then $M^{(\#)}\le a_0M_1^{(\#)}$ with a uniform constant $a_0$, and
hence if $M_1^{(\#)}<\infty$ then $M^{(\#)}<\infty$ for $\#=\text{G
or K}$, and all the results in Theorem \ref{thm:thm3.5} hold.
\end{prop}
\Proof The proof immediately follows from the definitions. In
particular, for the Kawasaki dynamics, we use the mean value theorem
and the boundedness of the partial derivatives for the function
$g_t(u,v)=((1+u)(1+v))^{-t}$.
\EndProof

In order to get the boundedness of $M_1^{(\#)}$, $\#=\text{G or
K}$, we have to know the quantities $\a(x;\x)-\a(x;u\x)$ in
\eqref{eq:eq3.17}-\eqref{eq:eq3.18} more in detail. For that purpose
we briefly introduce the restriction theory of reproducing kernel
Hilbert spaces \cite{A}. Let $\mcH$ be any RKHS (on $E$) with RK
$C(x,y)$. Let $R\sbs E$ be any (finite or infinite) subset of $E$,
and let $C_R(x,y)$, $x,y\in R$, denote the restriction of $C$ to the
set $R$. It was shown by Aronszajn that $C_R(x,y)$ is the RK of the
RKHS, call it $\mcH_{R,C_R}$, on the set $R$ consisting of all
restrictions of $\mcH$ to the set $R$ \cite{A}. We let $\|\cdot\|_C$
and $\|\cdot\|_{R,C_R}$ denote the norms on $\mcH$ and
$\mcH_{R,C_R}$, respectively. Then $\|\cdot\|_{R,C_R}$ is given by
for each $f\in \mcH_{R,C_R}$,
\begin{equation}\label{eq:eq3.15}
\|f\|_{R,C_R}:=\inf\{\|\wt{f}\|_C:\,\wt{f}(x)=f(x) \text{ for all
}x\in R\}.
\end{equation}
Moreover, for each $f\in \mcH_{R,C_R}$, there is a unique vector
$f'\in \mcH$ whose restriction to $R$ is $f$ and
\begin{equation}\label{eq:eq3.16}
\|f\|_{R,C_R}=\|f'\|_C.
\end{equation}
By the restriction theory for the RKHS's, the key terms
$\a(x;\x)-\a(x;u\x)$ in the above proposition have the following
representations (see \cite[Proposition 3.2]{Y3}):
\begin{prop}\label{prop:prop3.8}
For any $x\neq u\in E$
 and $x,u\notin \x\sbs E$, we have
 \begin{equation}\label{eq:eq3.19}
 \a(x;\x)-\a(x;u\x)=\left|(e_x,e_u)_{\x^c,B_{\x^c}}\right|^2\|e_u\|_{\x^c,B_{\x^c}}^{-2}.
 \end{equation}
In particular, in a formal way, we also have the representation:
\begin{equation}\label{eq:eq3.20}
 \a(x;\x)-\a(x;u\x)=\left|A(x,u)-A(x,\x)A(\x,\x)^{-1}A(\x,u)\right|^2\,\a(u;\x)^{-1}.
 \end{equation}
 \end{prop}
 In the next section we will discuss some sufficient conditions so
 that we can control $\a(x;\x)-\a(x;u\x)$ uniformly for $\x$.
\section{Examples}
In this section we discuss some examples for which the resulting
Glauber and Kawasaki dynamics are Fellerian Markov processes on
$C(\mcX)$ and leave invariant certain DPP's. In Proposition \ref{prop:prop3.7} we have seen that if we
could control the interdependencies; $M_1^{(\#)}<\infty$, then we are
done. Unfortunately we couldn't do it under our hypothesis (H), so
we impose further stronger conditions on the operator $A$. For any
complex number $z\in \mbC$, we let
$|z|_1:=|\text{Re}\,z|+|\text{Im}\,z|$.\\[2ex]
{\bf Assumption} (A): Let $A$ be a positive definite, bounded linear
operator on $\mcH_0\equiv l^2(E)$. We assume that there is a $\l>0$
such that
\[
\inf_{x\in E}\Big(A(x,x)-\sum_{y\neq x}|A(x,y)|_1\Big)\ge \l.
\]\\[1ex]
Any operator $A$ that satisfies the Assumption (A) is said to be diagonally dominant. 
When one considers the convolution operators on $l^2(\mbZ^d)$, it is
not hard to see that there are many operators $A$ that satisfy the
assumption (A). For example, let $C(x,y)\equiv C(x-y)$ be the
convolution operator
 on $l^2(\mbZ^d)$ coming from the Fourier coefficients $C(\cdot)$ of a sufficiently smooth positive function $\r(t)$
 on the torus $\mbT^d$. Then the components $C(x,y)$ decrease fast enough so that $\sum_{y}|C(x,y)|<\infty$. We may then take $A:=aI+C$ for
 some positive numbers $a$. 
 
Throughout
this section we suppose that our operators $A$ satisfy the
assumption (A). Let us fix a constant $q>0$ so that
\begin{equation}\label{eq:eq4.1}
q\ge q(A):=\sup_{x\in E}\sum_{y\neq x}|A(x,y)|_1.
\end{equation}
We define a $Q$-matrix $\wh{Q}$ on $E$ by
\begin{equation}\label{eq:eq4.2}
\wh{Q}(x,y)=\begin{cases}\frac1{q}|A(x,y)|_1,&x\neq y\\
-\sum_{y\neq x}\frac1q|A(x,y)|_1,&x=y.
\end{cases}
\end{equation}
Let $\wh{\P}$ be a stochastic matrix on $E$ defined by
\begin{equation}\label{eq:eq4.3}
\wh{\P}:=\wh{Q}+I.
\end{equation}
The following lemma is an analogue of \cite[Lemma 4.1]{SY}. The
proof follows by modifying the idea of the proof in \cite{SY} and we
leave it at the appendix for readers' convenience.
\begin{lem}\label{lem:lem4.1}
Assume that the operator $A$ satisfies the assumption (A). Then for
any $\x\sbs E$, the submatrix $A(\x,\x)$ is invertible and for any $x,y\in \x$,
\begin{equation}\label{eq:eq4.4}
|A(\x,\x)^{-1}(x,y)|\le M(x,y):=\begin{cases} \frac1\l,&x=y,\\\frac1\l \G(x,y),&x\neq y.\end{cases}
\end{equation}
where $\G:=\sum_{n=1}^\infty\left(\frac{q}{\l+q}\wh{\P}\right)^n$.
\end{lem}
With the help of the above lemma, we are able to state our main
result.
\begin{thm}\label{thm:thm4.2}
Suppose that the operator $A$ satisfies the assumption (A). Let us
take the flip rates $b(x;\x)$, $d(x;\x)$, and $c(x,y;\x)$ as stated
in Proposition \ref{prop:prop3.7}. Then the hypotheses of Theorem
\ref{thm:thm3.5} are satisfied and hence all the statements of
Theorem \ref{thm:thm3.5} hold and the DPP $\m$ corresponding to the
kernel operator $K:=A(I+A)^{-1}$ is invariant under the Glauber and
Kawasaki dynamics. Further, in addition to it, if $q(A)$ in
\eqref{eq:eq4.1} is sufficiently small, then the Markov processes are
ergodic and the statements in Theorem \ref{thm:thm3.6} hold.
\end{thm}
\Proof For the existence of Markov processes for Glauber and
Kawasaki dynamics, by Proposition \ref{prop:prop3.7}, it is enough
to check that $M_1^{(\#)}$, $\#=\text{G or K}$, defined in
\eqref{eq:eq3.17}-\eqref{eq:eq3.18}, are finite. Under the
assumption (A), \eqref{eq:eq3.20} has a rigorous meaning. By Lemma
\ref{lem:lem4.1}, $A$ and any submatrix of $A$ are boundedly invertible and we have
$\|A(\x,\x)^{-1}\|\le (\l+q)/\l^2$, uniformly for $\x\in E$. Then we easily see that
 $\a(x;\x)=\big(A(x\x,x\x)^{-1}(x,x)\big)^{-1}$, and Lemma \ref{lem:lem4.1} gives the bound:
\begin{equation}\label{eq:eq4.5}
\a(x;\x)\ge \l,\quad \text{uniformly for }x\in E \text{ and
}x\notin \x\sbs E.
\end{equation}
By using again the result of Lemma \ref{lem:lem4.1} in
\eqref{eq:eq3.20} we see that
\begin{eqnarray}\label{eq:eq4.6}
\sup_{\x\notni x,u}|A(x,\x)A(\x,\x)^{-1}A(\x,u)|&\le & \sup_{\x\notni
x,u}\sum_{y,z\in \x}|A(x,y)|M(y,z)|A(z,u)|\nonumber\\
&\le &\sum_{y,z\in E\sm\{x,u\}}|A(x,y)|M(y,z)|A(z,u)|.
\end{eqnarray}
Hence we have by \eqref{eq:eq3.20} and \eqref{eq:eq4.5}-\eqref{eq:eq4.6}
\begin{eqnarray}\label{eq:eq4.7}
M_1^{(\text{G})}&= &\sup_{x\in E}\sum_{u\neq x}\sup_{\x\notni
x,u}(\a(x;\x)-\a(x;u\x))\nonumber\\
&\le &\sup_{x\in E}\sum_{u\neq x}\Big(|A(x,u)|+\sum_{y,z\in
E\sm\{x,u\}}|A(x,y)|M(y,z)|A(z,u)|\Big)\frac{1}{\l}\nonumber\\
&\le
&\Big(q+q^2\frac{\l+q}{\l^2}\Big)\frac{1}{\l}=\frac{q}{\l}\Big(1+q\frac{\l+q}{\l^2}\Big)<\infty.
\end{eqnarray}
In a very similar way it follows that $M_1^{(\text{K})}<\infty$.

The flip or jump rates are chosen so that the Markov processes satisfy the detailed balance conditions w.r.t. the 
DPP $\m$. So, $\m$ is reversible for both Glauber and Kawasaki dynamics, and hence $\m$ 
is invariant under the dynamics (see \cite[Propositions 5.2 and 5.3, Chapter II]{Li}).

In order to check the ergodicity we have to know the quantities
$\e^{(\#)}$, $\#=\text{G or K}$. By using the definition of the flip and jump
rates and also by using the property \eqref{eq:eq4.5} it is not hard
to check that
\begin{equation}\label{eq:eq4.8}
\e^{(\text{G})}=1\quad \text{and} \quad \e^{(\text{K})}\ge
d_1\,\l\,\frac{1}{(1+\|A\|)^{2t}},
\end{equation}
where $d_1$ is the constant in \eqref{eq:eq3.8} and $\|A\|$ is the
operator norm of $A$. By Proposition \ref{prop:prop3.7}, \eqref{eq:eq4.7}, and \eqref{eq:eq4.8}, we see
that if $q$ is small enough, then $M^{(\#)}<\e^{(\#)}$, and this
completes the proof.
\EndProof
\renewcommand{\thesection}{\Alph{section}}
\setcounter{section}{0}
\section{Appendix}
In this appendix we provide with the proof of Lemma
\ref{lem:lem4.1}. The main ingredients of the method are the Markov
chain on the discrete set $E$, Feynman-Kac formula, and the
comparison of stochastic matrices of the Markov chains. The central idea
was introduced in the proof of \cite[Lemma 4.1]{SY}.\\[1ex]

\Proof[ of Lemma \ref{lem:lem4.1}] We first assume that $A$ is a
real matrix. For each $\x\sbs E$, we define a $Q$-matrix $Q_\x$ on
the set $\x\cup \ol{\x}\cup \{\ptl\}$, where $\ptl$ is an abstract
extra point playing as a cemetery (when $\x\equiv E$, we just ignore
it), and $\ol \x$ is also an abstract set consisting of the elements
of copies of $\x$; we denote them as $\ol \x:=\{\ol x:\,x \in \x\}$.
We define
\begin{eqnarray}\label{eq:eqA.1}
Q_\x(x,y)\equiv Q_\x(\ol x,\ol y)&:=&\frac1{q}A(x,y)_-,\quad x\neq
y\in
\x,\nonumber\\
Q_\x(x,\ol y)\equiv Q_\x(\ol x, y)&:=&\frac1{q}A(x,y)_+,\quad x\neq
y\in \x,\nonumber\\
Q_\x(x,\ol x)\equiv Q_\x(\ol x,x)&:=&0,\quad x\in \x,\nonumber\\
Q_\x(x, x)\equiv Q_\x(\ol x,\ol x)&:=&-\sum_{y\in E;\,y\neq x}\frac1{q}|A(x,y)|,\quad x\in \x,\nonumber\\
Q_\x(x, \ptl)\equiv Q_\x(\ol x,\ptl)&:=&\sum_{y\in E\sm \x}\frac1{q}|A(x,y)|,\quad x\in \x,\nonumber\\
Q_\x(\ptl, x)\equiv Q_\x(\ptl,\ol x)&\equiv&
Q_\x(\ptl,\ptl):=0,\quad x\in \x,
\end{eqnarray}
here $a_\pm:=\max\{\pm a,0\}$ for each real number $a$. Given a
function $f\in l^\infty(\x)$, we extend it to a function $\wt f\in
l^\infty(\x\cup\ol \x\cup \{\ptl\})$ by
\begin{equation}\label{eq:eqA.2}
\wt f(\wt x):=\begin{cases}f(x),&\wt x=x\\
-f(x),&\wt x=\ol x\\
0,&\wt x=\ptl.
\end{cases}
\end{equation}
We consider the anti-symmetric subspace $l_a^\infty(\x\cup\ol \x\cup
\{\ptl\})$ of $l^\infty(\x\cup\ol \x\cup \{\ptl\})$ defined by
\begin{equation}\label{eq:eqA.3}
l_a^\infty(\x\cup\ol \x\cup \{\ptl\}):=\{f\in l^\infty(\x\cup\ol
\x\cup \{\ptl\}):\,f(\ol x)=-f(x), \,\,f(\ptl)=0\}.
\end{equation}
We notice that $Q_\x$ can be regarded as an operator on
$l_a^\infty(\x\cup\ol \x\cup \{\ptl\})$. We define a function
$V_\x(\wt x)$ on $\x\cup\ol \x\cup \{\ptl\}$ by for $x\in \x$,
\begin{equation}\label{eq:eqA.4}
V_\x(x)=V_\x(\ol x):=\frac1{q}\Big(-A(x,x)+\sum_{y\in E:\,y\neq
x}|A(x,y)|\Big),\text{ and }\quad V_\x(\ptl)=0.
\end{equation}

Now we consider the equation:
\begin{equation}\label{eq:eqA.5}
(A(\x,\x)f)(x)=h(x),\quad x\in \x,
\end{equation}
for $h\in l^\infty(\x)$. By the definition of the $Q$-matrix $Q_\x$,
the equation \eqref{eq:eqA.5} is equivalent to
\begin{equation}\label{eq:eqA.6}
-(Q_\x+V_\x)\wt f=\frac1{q}\wt h.
\end{equation}
Let us define a stochastic matrix $\P_\x$ on $\x\cup\ol \x\cup
\{\ptl\}$ by
\begin{equation}\label{eq:eqA.7}
\P_\x:=Q_\x+I_{\x\cup\ol \x\cup \{\ptl\}}.
\end{equation}
Let $\{X_t^{(\x)}:\,t\ge 0\}$ be the Markov chain on $\x\cup\ol
\x\cup \{\ptl\}$ generated by $Q_\x$. Then by the Feynman-Kac
formula, we have
\begin{eqnarray}\label{eq:eqA.8}
\wt f(\wt x)&=&-\frac1{q}(Q_\x+V_\x)^{-1}\wt h(\wt x)\nonumber\\
&=&\frac1q \mbE_{\wt x}\left[\int_0^\infty\wt
h(X_t^{(\x)})\exp\left(\int_0^tV_\x(X_s^{(\x)})ds\right)dt\right].
\end{eqnarray}
Let us now take $h:=\d_y$, the Dirac function at the point $y$. Then
the solution $f(x)$ in \eqref{eq:eqA.5} is $f(x)=A(\x,\x)^{-1}(x,y)$,
thus we have (notice that $\wt {\d_y}=\d_y-\d_{\ol y}$)
\begin{equation}\label{eq:eqA.9}
|A(\x,\x)^{-1}(x,y)|\le
\frac1q\mbE_{x}\left[\int_0^\infty\d_y(\p(X_t^{(\x)}))\exp\left(\int_0^tV_\x(\p(X_s^{(\x)}))ds\right)dt\right],
\end{equation}
where $\p$ is a projection operator on $\x\cup\ol \x\cup \{\ptl\}$
defined by
\begin{equation}\label{eq:eqA.18}
\p(x)\equiv \p(\ol x):=x\quad \text{and}\quad \p(\ptl):=\ptl.
\end{equation}
We would like to estimate the r.h.s. of \eqref{eq:eqA.9}. For this,
we introduce another $Q$-matrix $\wh Q_\x$ on $\x\cup \{\ptl\}$ by
\begin{eqnarray}\label{eq:eqA.10}
\wh Q_\x(x,y)&=&\begin{cases}\frac1q|A(x,y)|,&x\neq y,\,\,x,y\in
\x,\\
-\frac1q\sum_{z\in E:\,z\neq x}|A(x,z)|,&x=y\in \x,
\end{cases}\nonumber\\
\wh Q_\x(x,\ptl)&=& \frac1q\sum_{z\in E\sm \x}|A(x,z)|,\quad x\in
\x,\nonumber\\
\wh Q_\x(\ptl,y)&\equiv&\wh Q_\x(\ptl,\ptl)=0,\quad y\in \x.
\end{eqnarray}
We also define a stochastic matrix $\wh \P_\x$ on $\x\cup \{\ptl\}$
by
\begin{equation}\label{eq:eqA.12}
\wh \P_\x:=\wh Q_\x+I_{\x\cup \{\ptl\}}.
\end{equation}
Then we notice that the probability law of the chain
$\{\p(X_t^{(\x)})\}$ on $\x\cup \{\ptl\}$ is the same as that of the
Markov chain on $\x\cup \{\ptl\}$ generated by $\wh Q_\x$. By the
assumption (A) we notice that $V_\x\le -\l/q$. Thus by using the
strong Markov property we get
\begin{equation}\label{eq:eqA.14}
|A(\x,\x)^{-1}(x,y)|\le \frac1q\|(\wh
Q_\x+V_\x)^{-1}\|_\infty\,\mbE_x\left[e^{-\l/q\t_y^{(\x)}};\,\t_y^{(\x)}<\infty\right],
\end{equation}
where $\t_y^{(\x)}$ is the first hitting time at $y$ of the chain
$\{\p(X_t^{(\x)})\}$, and $\|(\wh Q_\x+V_\x)^{-1}\|_\infty$ is the operator
norm of $(\wh Q_\x+V_\x)^{-1}$ acting on $l^\infty(\x\cup\{\ptl\})$,
which satisfies the bound:
\begin{equation}\label{eq:eqA.15}
\|(\wh Q_\x+V_\x)^{-1}\|_\infty\le q/\l.
\end{equation}
Since $\t_x^{(\x)}=0$ for the Markov process starting at $x$, we see from \eqref{eq:eqA.14} and \eqref{eq:eqA.15} that
\begin{equation}\label{eq:eqA.15-1}
|A(\x,\x)^{-1}(x,x)|\le 1/\l.
\end{equation}
This is the bound in the lemma for the diagonal components. In order to estimate $|A(\x,\x)^{-1}(x,y)|$ for $x\neq y$, we define
\begin{equation}\label{eq:eqA.16}
u_y^{(\x)}(x):=\mbE_x\left[e^{-\l/q\t_y^{(\x)}};\,\t_y^{(\x)}<\infty\right], \quad x\neq y.
\end{equation}
If we let $\t$ the random variable which is exponentially
distributed with parameter $1$, then we have the identity:
\begin{eqnarray}\label{eq:eqA.17}
u_y^{(\x)}(x)&=&\mbE\left[e^{-\l/q\t}\right]\Big[\wh
\P_\x(x,y)+\sum_{z\in \x:\,z\neq y}\wh
\P_\x(x,z)u_y^{(\x)}(z)\Big]\nonumber\\
&=&\Big(\frac{q}{\l+q}\Big)\Big[\wh \P_\x(x,y)+\sum_{z\in
\x:\,z\neq y}\wh
\P_\x(x,z)u_y^{(\x)}(z)\Big]\nonumber\\
&=&\sum_{n=1}^\infty\Big(\frac{q}{\l+q}\wh \P_\x\Big)^n(x,y).
\end{eqnarray}
Now by the definition of $\wh Q_\x(x,y)$ in \eqref{eq:eqA.10} we see
that for $\x\sbs \x'$,
\begin{equation}\label{eq:eqA.18}
\wh \P_\x(x,y)=\wh \P_{\x'}(x,y)\quad\text{whenever }x,y\in \x.
\end{equation}
Also, since $\ptl$ plays as a cemetery, once the process visits
$\ptl$, it never comes out from it. Therefore, the nonzero
contributions in the term $\big(\wh \P_\x\big)^n(x,y)$ come only
from the random walk path of length $n$ connecting $x$ and $y$ on
the set $\x$ (avoiding the cemetery $\ptl$). Obviously, such a number of
paths increases as the set $\x$ increases. Therefore the last presentation of
\eqref{eq:eqA.17} is bounded by $\G(x,y)$ where
\begin{equation}\label{eq:eqA.19}
\G:=\sum_{n=1}^\infty\left(\frac{q}{\l+q}\wh \P\right)^n,
\end{equation}
with $\wh \P:=\wh \P_E$. Inserting this and \eqref{eq:eqA.15} into
\eqref{eq:eqA.14} we get the bound in the lemma for off-diagonal
components. Together with \eqref{eq:eqA.15-1} we are done in the
case that $A$ is a real matrix. When $A$ is a complex matrix, we
write $A=A_1+iA_2$, where $A_1$ and $A_2$ are real matrices. Let
$E_1$ and $E_2$ be two copies of $E$. Then we have the bijection
\begin{equation}\label{eq:eqA.20}
l^2(E)\ni f=f_1+if_2\mapsto f_1\oplus f_2\in
l^2_{\text{real}}(E_1)\oplus l^2_{\text{real}}(E_2)\cong
l^2_{\text{real}}(E_1\cup E_2),
\end{equation}
where $l^2_{\text{real}}(\cdot)$ means the real Hilbert space. Under
this map, $A$ in $l^2(E)$ is equivalent to the matrix
\begin{equation}\label{eq:eqA.21}
\wt A=\left(\begin{matrix}A_1&-A_2\\
A_2&A_1\end{matrix}\right)
\end{equation}
acting on $l^2_{\text{real}}(E_1\cup E_2)$. Similary, for any subset
$\x\sbs E$, the submatrix $A(\x,\x)$ acting on $l^2(\x)$ is
equivalent to the submatrix
\begin{equation}\label{eq:eqA.22}
\wt A(\x,\x)=\left(\begin{matrix}A_1(\x,\x)&-A_2(\x,\x)\\
A_2(\x,\x)&A_1(\x,\x)\end{matrix}\right)
\end{equation}
of $\wt A$ acting on $l^2_{\text{real}}(\x_1\cup \x_2)$, where
$\x_i$, $i=1,2$, are again the copies of $\x$. Notice that the
enlarged real matrix $\wt A$ satisfies the conditions of the lemma
with $E$ being replaced by $E_1\cup E_2$. Let $A(\x,\x)^{-1}\equiv
C+iD$, where $C$ and $D$ are real matrices. Then we can check that
\begin{equation}\label{eq:eqA.23}
C=P_{\x_1}\wt A(\x,\x)^{-1}P_{\x_1}\quad \text{and}\quad
D=-P_{\x_1}\wt A(\x,\x)^{-1}P_{\x_2},
\end{equation}
where $P_{\x_i}$, $i=1,2$, are the projections on
$l^2_{\text{real}}(\x_1\cup \x_2)$ onto $l^2_{\text{real}}(\x_i)$,
$i=1,2$, respectively. So, $|A(\x,\x)^{-1}(x,y)|\le
|C(x,y)|+|D(x,y)|$, and by \eqref{eq:eqA.23} and applying the result
for the real case we also arrive at the conclusion for the complex
case. \EndProof

\bigskip
 \noindent{\it Acknowledgments.} We thank the referee for valuable comments which improved the paper greatly.
 This work was supported by the Korea--Japan Basic Scientific
 Cooperation Program "Noncommutative Stochastic Analysis and Its
 Applications to Network Science".

\end{document}